\documentclass[journal,twocolumn]{IEEEtran}
\IEEEoverridecommandlockouts
\usepackage{latexsym,amssymb,setspace,array,multirow}
\usepackage{makecell}
\usepackage{algorithm}
\usepackage{algorithmicx}
\usepackage{algpseudocode}
\usepackage{amsmath}
\usepackage{subfigure}
\usepackage[dvips]{graphicx}
\usepackage{booktabs}
\usepackage{cases}
\usepackage{color}
\usepackage{cite}
\usepackage{mathrsfs}
\usepackage{amsbsy}
\usepackage{comment}
\usepackage{ulem}
\usepackage[colorlinks,linkcolor=red,anchorcolor=red,citecolor=red]{hyperref}
\definecolor{pink}{rgb}{0.58,0,0.83}
\definecolor{orange}{rgb}{1,0.5,0}
\definecolor{lightgreen}{rgb}{0.2, 0.8, 0.2}
\definecolor{lightyellow}{rgb}{0.84, 0.65, 0.13}

\graphicspath{ {./images1/}}
\usepackage{CJK}  

\newtheorem{remark}{Remark}

\newcommand{\degree}{^\circ}
\renewcommand{\theenumi}{\roman{enumi}}

\newcounter{step}
\renewcommand{\thestep}{S\arabic{step}}

\newcounter{assum}
\renewcommand{\theassum}{A\arabic{assum})}

\newcommand{\skipthis}[1]{ }

\usepackage{color}
\definecolor{red}{rgb}{1,0,0}
\definecolor{black}{rgb}{0,0,0}

\begin{document}
  \title{DDK: A Deep Koopman Approach for Dynamics Modeling and Trajectory Tracking of Autonomous Vehicles}
  \providecommand{\keywords}[1]{\textbf{\textit{Index terms---}} #1}
  \author{Yongqian Xiao
    \thanks{}
     } \maketitle
\begin{abstract}
	Autonomous driving has attracted lots of attention in recent years. An accurate vehicle dynamics is important for autonomous driving techniques, e.g. trajectory prediction, motion planning, and control of trajectory tracking. Although previous works have made some results, the strong nonlinearity, precision, and interpretability of dynamics for autonomous vehicles are open problems worth being studied. In this paper, the approach based on the Koopman operator named deep direct Koopman (DDK) is proposed to identify the model of the autonomous vehicle and the identified model is a linear time-invariant (LTI) version, which is convenient for motion planning and controller design. In the approach, the Koopman eigenvalues and system matrix are considered as trainable tensors with the original states of the autonomous vehicle being concatenated to a part of the Koopman eigenfunctions so that a physically interpretable subsystem can be extracted from the identified latent dynamics. Subsequently, the process of the identification model is trained under the proposed method based on the dataset which consists of about 60km of data collected with a real electric SUV while the effectiveness of the identified model is validated. Meanwhile, a high-fidelity vehicle dynamics is identified in CarSim with DDK, and then, a linear model predictive control (MPC) called DDK-MPC integrating DDK is designed to validate the performance for the control of trajectory tracking.  Simulation results illustrate that the model of the nonlinear vehicle dynamics can be identified effectively via the proposed method and that excellent tracking performance can be obtained with the identified model under DDK-MPC.
\end{abstract}
\section{Introduction}   \label{dfahc:sec:introduction}
\IEEEPARstart Autonomous driving technologies have been developed in recent decades because of their promising economy, military, and societal values. Vehicle dynamics modeling and control are two important parts of autonomous driving. For vehicle dynamics, classic kinematics models are usually adopted for low-speed scenes\cite{tu2013dynamic}. On the contrary, dynamics models are suitable for high-speed scenarios and even limited conditions, such as drifting\cite{goh2020toward}. Time-varying and nonlinear vehicle dynamics are too complicated to obtain precise approximation models so that some decent analytical models are established, e.g. 14-DOF dynamic model\cite{zheng2013decision}, separated longitudinal or lateral dynamics\cite{rajamani2011vehicle}, Pacejka tire model\cite{hewing2018cautious}, and linear time varying dynamics \cite{falcone2008linear}. \par
Notably, in \cite{vicente2020linear}, it was discovered that a linear data-driven dynamics was adequate to approximate a second order vehicle dynamics, but it only identified the longitudinal velocity and yaw rate, and did not validate with control experiments. End-to-end (E2E) methods based on deep learning are also an important category of identification method for vehicle dynamics. Recurrent neural network was adopted to estimate side-slip angle for a simplified analytic kinematics\cite{graber2018hybrid}. Spielberg et  al.\cite{spielberg2019neural} constructed a vehicle dynamics with a multi-layer perception (MLP) that took yaw rate, longitudinal and lateral velocities, steering angle, and front longitudinal force as state, and the yaw and lateral acceleration as input. Da Lio\cite{da2020modelling} has studied the influences of different neural network structures on the identification performances of the longitudinal vehicle dynamics. However, the obtained systems by the above E2E-based methods are generally nonlinear dynamics and lack interpretability.
 \par
As a modeling approach, the Koopman operator has received lots of attention since it is a linear operator which has the ability to approximate nonlinear dynamics in some infinite-dimensional spaces \cite{koopman1931hamiltonian}. The Koopman operator is easy to be applied with its linearity and convergence while it is difficult to deal with due to its infinite dimension. Many approaches have been proposed and developed to approximate the Koopman operator in an acceptable dimension. Dynamic mode decomposition (DMD) \cite{schmid2010dynamic} and extended DMD (EDMD) \cite{williams2015data} are two widely used methods based on singular value decomposition (SVD) and least square (LS) separately. Besides, variants were proposed, such as bilinear Koopman\cite{bruder2021advantages}. DMD and EDMD are advanced to be compatible with forced systems\cite{proctor2018generalizing}\cite{williams2016extending}. EDMD and kernel-based DMD\cite{kevrekidis2016kernel} have better performance than DMD because they can manually design kernel functions for feature extraction before decomposition. Nevertheless, kernel functions critically decide the approximating effectiveness and appropriate kernel functions are arduous to discover especially for complicated nonlinear systems. Neural networks are congenitally suitable for taking the place of kernel functions, and considerable related works have been proposed,  
e.g., Deep Koopman\cite{lusch2018deep} for learning time-invariant Koopman eigenvalues, variation-based Koopman thoughts, dictionary (kernel) functions learning \cite{li_extended_2017}\cite{otto2019linearly}, etc. For vehicle dynamics, there are existing two works based on EDMD for identifying vehicle dynamics, i.e., a bicycle model based on EDMD\cite{cibulka2019data}, and a four-wheel dynamics based on DeepEDMD\cite{xiao2020deep}.
  
Inspired by LRAN\cite{otto2019linearly}, DeepEDMD\cite{xiao2020deep}, and Deep Koopman\cite{lusch2018deep}, this paper develop an approach called deep direct Koopman (DDK) that the `direct' reflects in two aspects. One is that DDK directly deals with the Koopman eigenvalues and system matrix as trainable tensors so that the identified model is described in LTI form with a diagonal state transition matrix, and it is suitable for forced dynamics. The other is that DDK directly concatenates the original state to be a part of the Koopman eigenfunctions so that an interpretable subsystem can be extracted from the identified model. Concatenating the original state is not feasible for systems that take raw pixels as inputs \cite{van2020deepkoco}\cite{xiao2021cknet} since the ground truth of states is inaccessible.
 \par
 Vehicle trajectory tracking control is also a critical part of autonomous driving for tracking results from motion planning module. There are many control algorithms for vehicle control, such as controllers based on proportion integration differentiation (PID) and pure pursuit\cite{samuel2016review}, MPC\cite{hewing2018cautious}\cite{xiao2020deep}, learning-based MPC\cite{ostafew2016learning}, linear quadratic regulator (LQR)\cite{chen2019autonomous}, and E2E control approaches including supervised learning and deep reinforcement learning (DRL)\cite{hsu2018end}\cite{kendall_learning_2019}. PID is the most commonly used in practice because it is model-free and stable. MPC can acquire more precise control performance, but it is limited to use due to the strong uncertainty and nonlinearity of vehicle dynamics. 
 
 Overall, the existing control algorithms of vehicles meet the requirements for all most scenarios, e.g. urban roads, freeway, parking, and et al. This paper validates the control performance in a point-to-point (P2P) way for some special scenarios, i.e. multi-vehicle formation control with specified geometry, overtaking or lane change maneuvers in dense traffic flows, and controlling multi-vehicle to pass through intersections without traffic lights under no waiting\cite{li2021optimal}. 
 P2P trajectory tracking demands vehicles to track the reference positions, yaw angles, longitudinal and lateral velocities, and yaw rate corresponding to the specific time sequence. It is expectable that vehicles in cities will be controlled uniformly to improve traffic efficiency so that the aforementioned scenarios will be common and P2P trajectory tracking will be necessary. To sum up, the main contributions of this work are as follows.

  \par
	\begin{itemize}
		\item A deep learning approach called DDK is developed for approximating the Koopman operator of vehicle dynamics by directly learning the Koopman eigenvalues and input matrix. The identified model is LTI and an interpretable subsystem can be extracted from the identified latent dynamics. DDK is utilized to identify the dynamics of a real electric SUV and a sedan car of CarSim.
		\item A linear MPC called DDK-MPC is proposed based on the predictive model which is identified by DDK. Simulation results validate the feasibility of DDK-MPC for P2P trajectory tracking under the condition of real-time requirements. 
	\end{itemize}

	The rest of this paper is arranged as follows. Section \ref{sec:DDK} introduces the Koopman theory of vehicle dynamics, details of DDK, and related loss functions. In Section \ref{sec:design_MPC}, DDK-MPC is designed to realize P2P trajectory tracking, and simulations on prediction and P2P trajectory tracking are conducted In Section \ref{sec:simulations} and followed by Section \ref{sec:experiment} that prediction experiments on datasets collected by a real vehicle are demonstrated. Finally, conclusions and future research directions are drawn in Section \ref{sec:conclusion}.

\section{DDK for modeling vehicle dynamics}\label{sec:DDK}
In this work, the four-wheel vehicle dynamics the same as the simulation model in Carsim is utilized to carry out the simulation and analysis \cite{xiao2020deep}, which is given as follows.
\begin{equation}\label{equ:non_linear_vehicle_dynamics}
	f\left( s_{t+1} \right) =f\left( s_t,u_t \right) 
\end{equation}
where $s \in \mathbb{R}^{m}$ denotes the vehicle state vector consisting of poses $(x, y, \psi)$, and the assicoated velocities $(v_{x}, v_{y}, \dot{\psi})$. Poses include the longitudinal and lateral position, and yaw angle in the vehicle coordinate, and the associated velocities including the longitudinal and lateral velocities, and yaw rate. $u_t \in \mathbb{R}^n$ is the control vector including the steering wheel angle and the engine, where the engine consists of the throttle opening and brake pressure. \par

\begin{figure*}[htbp]
		\setlength{\abovecaptionskip}{-0.1cm}
		\setlength{\belowcaptionskip}{-1cm}
		\begin{center}
			\includegraphics[scale=0.75]{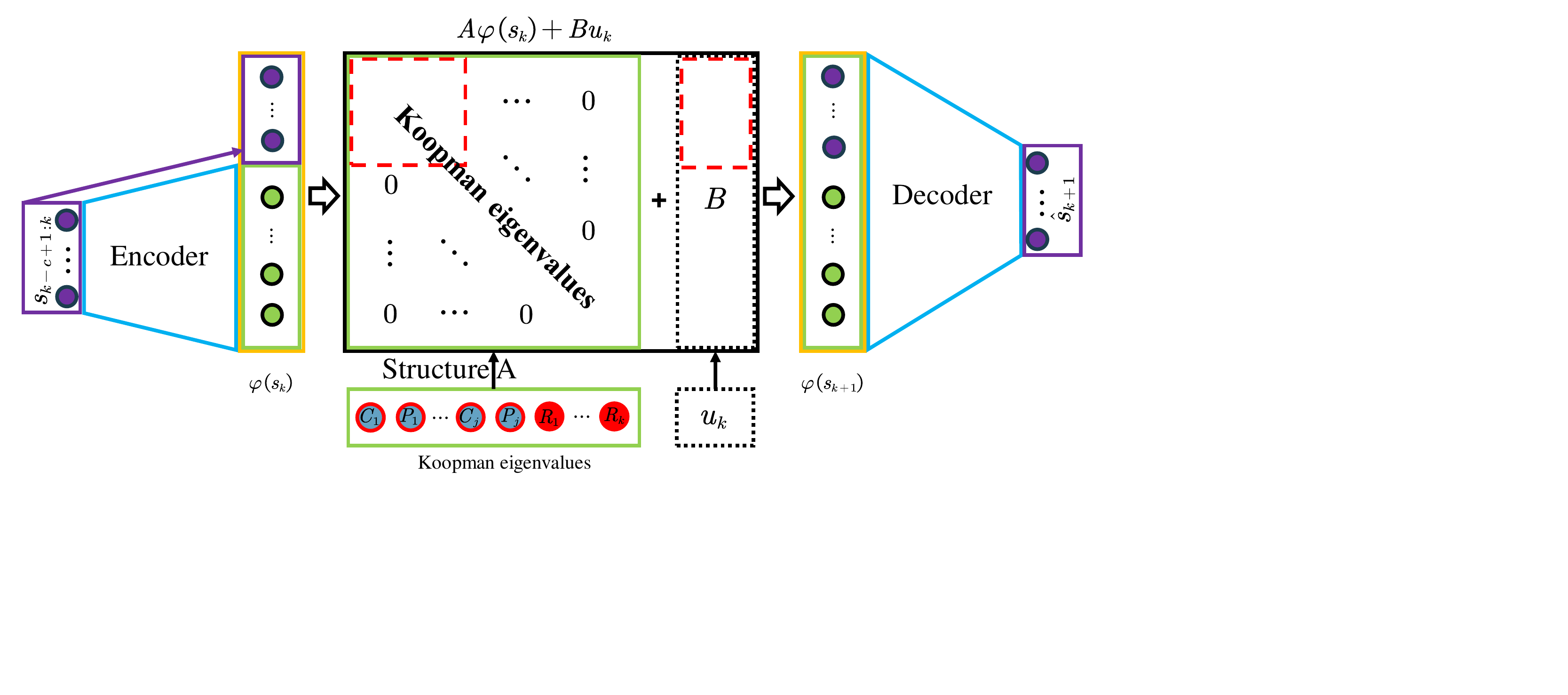}
			\caption{The neural network framework of DDK. This framework adopts the structure of anto-encoder (AE), and the original state is concatenated to the outputted features by the encoder. $\varphi(s_k)$ is an approximation of the Koopman eigenfunctions, and it is also the latent state of the approximated linear vehicle dynamics. Especially, the state transition matrix is constructed with the learned Koopman eigenvalues via structuring Jordan block. Red dashed wireframes express the interpretable subsystem described by the $m$-th leading principal submatrix of $A$ and $B$. This framework is also suitable for unforced dynamics while $u_k$ equals $\boldsymbol{0}_{n \times 1}$ constantly. $R_k$ denotes the $k$-th real eigenvalue while $C_j$, $P_j$ denote the real and imaginary parts of the $j$-th conjugate complex pair respectively, and $c$ indicates that adjacent $c$ frames of vehicle states are concatenated as a new state.}
			\label{fig:framework}
		\end{center}
	\end{figure*}

	\textbf{Construction of DDK: }
	The neural network framework of DDK is depicted in Fig. \ref{fig:framework}. DDK concatenates the original state with the output of the encoder and directly copes with the Koopman eigenvalues as trainable tensors to parameterize Jordan blocks for establishing diagonal state transition matrix $A$ for discrete state-space systems.
 	For discrete-time vehicle dynamics of \eqref{equ:non_linear_vehicle_dynamics}, the Koopman operator can be described as follows.
	\begin{equation}
		\varphi(s_{k+1}) = \mathcal{K} \varphi(s_k) = \varphi(f(s_k, u_k))
	\end{equation}
	where $\mathcal{K}$ is the Koopman operator with the property of the linearity in Hilbert space $\mathcal{H}$ spanned by the Koopman eigenfunctions $\varphi$. The Koopman operator identifies the nonlinear system $f$ in a linear form even though in some infinite spaces. For each Koopman eigenfunction, the Koopman operator can evolve linearly with the associated Koopman eigenvalue\cite{parmar2020survey}.
	\begin{equation}
 	\mathcal{K} \varphi_i = \mu_i \varphi_i
	\end{equation}
	where $\mu_i$ is the Koopman eigenvalue associated with the eigenfunction $\varphi_i$. It can be extended to $p$ steps operator $\mathcal{K}^p$ because of its linearity, which is $\mathcal{K}^p \varphi_i = \mu_i^p \varphi_i$. To handle with the intractability of infinity, the Koopman eigenfunctions are approximated with features consisting of the original state followed by characteristics extracted by the encoder.
	\begin{equation}\label{eq:koopman_eigenfunctions}
		\varphi(s_k) = [s_k; \phi_e(s_k, \theta_e)]
	\end{equation}
	where $\varphi(s_k) \in \mathbb{R}^K$ is also the latent state. $\phi_e$ denotes the encoder parameterized with $\theta_e$. 
	\begin{remark}
		Concatenating $s_k$ and $\phi_e(s_k, \theta_e)$ has three advantages. Firstly, all critical features can be made sure to be contained in the latent state. Secondly, the higher interpretability of the approach can be obtained since the former $m$ elements of $\varphi(s_k)$ match the physical vehicle state. Finally, optimization problems about trajectory tracking could be built mainly focusing on the former $m$ elements.
	\end{remark}

	DDK parameterizes the state transition matrix by learning the Koopman eigenvalues directly. For situations only with real eigenvalues, $\Lambda = [R_1, \cdots, R_K]$, the transition matrix can be constructed by $A=diag([R_1, R_2, \cdots, R_3])$, where $R_i$ denotes the $i$-th eigenvalue.
	 For situations also with conjugate complex pairs, which is $\Lambda = [C_1, P_1, \cdots, C_j, P_j, R_1, \cdots, R_k]$. $A$ can be constructed to a block-diagonal matrix.
	\begin{equation}\label{eq:A}
		A = diag([\mathcal{B} _1, \cdots, \mathcal{B} _j, R_1, \cdots ,R_k])
	\end{equation}
	with $\mathcal{B}_j$ is $2\times 2$ matrix corresponding to $i$-th conjugate complex pairs.
	\begin{equation}\label{eq:jordan_block}
		\mathcal{B} _j=\left[ \begin{matrix}
			C_j&		P_j\\
			-P_j&		C_j\\
		\end{matrix} \right] 
	\end{equation}
	where $C_i, P_j$ denote the real and imaginary part of the $j$-th conjugate complex pair of eigenvalues respectively with $K=2j+k$. $A$ is the $K$-order approximation of the Koopman operator. Actually, $A$ also can be constructed as Jordan Canonical Form \cite{mezic2017koopman}. For forced dynamics, the Koopman operator evolves as follows.
	\begin{equation}\label{eq:decoupled_dynamics}
		\varphi(s_{k+1}) = \mathcal{K}\varphi(s_k) \doteq A\varphi(s_k) + Bu_k
	\end{equation}

	Naturally, it can be extended to a $p$ steps evolution:
	\begin{equation}\label{eq:multi-step-KO}
		\varphi(s_{k+p})= \mathcal{K}^p\varphi(s_k) \doteq A^p\varphi \left( s_k \right) +\sum_{i=1}^p{A^{i-1}Bu_{k+p-i}}
	\end{equation}

	The Koopman modes remap latent states back to the original state space, which can be obtained by solving a least square (LS) problem in EDMD. Neural networks in lieu of the Koopman modes can improve remapping performance\cite{otto2019linearly}.
	\begin{equation}\label{eq:decoder}
		\hat{s}_k = \phi_d(\varphi(s_k), \theta_d)
	\end{equation}
	where $\phi_d$ denotes the decoder parameterized by $\theta_d$. \par	
	
	To sum up, the approximated latent dynamics can be described by the following LTI system.
	\begin{equation}\label{eq:final_dynamics}
		\begin{cases}
			\varphi \left( s_{k+1} \right) =A\varphi \left( s_k \right) +Bu_k\\
			\hat{s}_k=\phi _d\left( \varphi \left( s_k \right) \right)\\
		\end{cases}
	\end{equation}
	\textbf{Interpretability analysis: }
	The state transition matrix is diagonally constructed with \eqref{eq:A}. According to \eqref{eq:koopman_eigenfunctions} and \eqref{eq:multi-step-KO}, it is clear that the former $m$ elements of Koopman eigenfunctions match the physical vehicle state no matter how many evolution steps have been done. As a result, a subsystem can be extracted from \eqref{eq:decoupled_dynamics} shown in Fig. \ref{fig:framework} with red dashed frames.
	\begin{equation}\label{eq:interpretable_dynamics}
		s_{k+1} = A_m s_k + B_m u_k
	\end{equation}
	where $A_m$ and $B_m$ are the $m$-th leading principal submatrix of $A$ and $B$ respectively. The state and control corresponds to the original vehicle dynamics in \eqref{equ:non_linear_vehicle_dynamics}. In this way, this subsystem is regarded as an interpretable approximation.\par
	
	\textbf{Loss Functions: }In this work, we adopt the same loss functions as Deep Koopman\cite{lusch2018deep}, LRAN\cite{otto2019linearly}, and DeepEDMD\cite{xiao2020deep}. These loss functions include reconstruction loss, linear loss, multi-step reconstruction loss, and regularization loss. They constrain the reconstruction, linearity, multi-step prediction performances, and avoid over-fitting respectively.
	\begin{equation}\label{equ:loss_function}
		\begin{aligned}
			\mathcal{L} _r &= \frac{1}{p}\sum_{i=1}^p{\left\| s_i-\phi _d [ s_i;\phi _e( s_i ) ] \right\| _{2}^{2}} \\
			\mathcal{L} _l &= \frac{1}{p}\sum_{i=1}^p{\left\| \left[ s_{i};\phi _e\left( s_{i} \right) \right] -\mathcal{K} ^i[s_0; \phi_e \left( s_0 \right)] \right\| _{2}^{2}} \\
			\mathcal{L} _{mr} &= \sum_{i=1}^p{\left\| s_{i}-\phi _d\left( \mathcal{K} ^i\varphi \left( s_0 \right) \right) \right\| _{2}^{2}}\\
			l_2&=\left\| \theta _e \right\| _{2}^{2}+\left\| \theta _d \right\| _{2}^{2}+\left\| \varLambda \right\| _{2}^{2}+\left\| B \right\| _{2}^{2}
		\end{aligned}
	\end{equation}

	The approximation process of the Koopman operator is built as an optimization problem to minimize the following weighted loss.
	\begin{equation}\label{eq:weighted_loss}
		\mathcal{L} = \alpha_1 \mathcal{L} _r + \alpha_2 \mathcal{L} _l + \alpha_3 \mathcal{L} _{mr} + \alpha_4 l_2
	\end{equation}
	where $\alpha _i$ is the corresponding weight to different losses that expresses the importance of each loss. Implementation steps are outlined in algorithm \ref{alg:DDK}.

\begin{algorithm}
	\caption{Implementation steps for DDK}\label{alg:DDK}
	\begin{algorithmic}[1]
		\State Initialize ${\theta}_e$, ${\theta}_d$, $\Lambda$, $B$, time step $p$, $\alpha_i$, $i=1,\cdots,4$, batch size $b_s$, learning rate $\beta$
		\Repeat
		\State Sample a batch of sequences and transform them to randomly selected vehicle coordinates.
		\State Obtain the Koopman eigenfunctions $\varphi(s_{0:p})$ with \eqref{eq:koopman_eigenfunctions} and reconstruction states $\hat{s}_{0:p}$ with \eqref{eq:decoder}.
		\State Structure $A$ according to $\Lambda$ with \eqref{eq:A}.
		\State Perform $p$ steps Koopman operator with \eqref{eq:multi-step-KO} and reconstruct them with \eqref{eq:decoder} to get $\varphi(s_{1:p})$ and $\phi_d(\varphi_e({s}_{1:p}))$.
		\State Calculate the weigted loss with \eqref{eq:weighted_loss}.
		\State Update $\theta_e$, $\theta_d$, $\Lambda$, $B$ with an Adam optimizer.
		\Until The epoch terminated
	\end{algorithmic}
\end{algorithm}

\section{DDK-MPC for P2P trajectory tracking}\label{sec:design_MPC}
	In this section, DDK-MPC is designed for P2P trajectory tracking. P2P trajectory tracking demands the vehicle to follow the reference trajectory including poses and associated velocities along the time sequence. As a result, the P2P trajectory tracking task can be described as the following optimization problem:
	\begin{equation}\label{eq:P2P_problem}
		\min_{u_{1:T}} \sum_{k}{(\left\| s_k-s_{k}^{r} \right\| _1+\left\| \Delta  u_k \right\| _1)}
	\end{equation}
where $s^r_k$ denotes the reference state at $k$. $\Delta u_k$ is the control increment at $k$ to improve smoothness of tracking. Note that P2P tracking does not reset the reference trajectory according to the nearest point.
	The approximated latent dynamics \eqref{eq:final_dynamics} can be represented as the following incremental form.
	\begin{equation}\label{eq:MPC_dynamics}
		\begin{aligned}
			\varPhi(s_{k+1}) &= \mathbb{A} \varPhi (s_k) + \mathbb{B} \Delta u_k	\\
			y_{k} &= \mathbb{C} \varPhi(s_{k})
		\end{aligned}
	\end{equation} 
	where $\Delta u_k = u_k - u_{k-1}$ is the control increment at $k$, $y_{k}$ denotes the system's output at $k$ and 
	\begin{equation}\nonumber
		\begin{aligned}
			&\varPhi (s_k)=\left[ 
				\varphi ^{\top}(s_k),	\;\;	u_{k-1}^{\top}
			\right] ^{\top},\; \mathbb{A} =\left[ \begin{matrix}
				A&		B\\
				\boldsymbol{0}_{n\times K}&		I_n\\
			\end{matrix} \right] 
			\\
			& \mathbb{B} =[ B^{\top},\; I_{n}^{\top}] ^{\top},\;\; \mathbb{C} =[\mathrm{diag}([I_m, \boldsymbol{0}_{K-m}]),\;	\boldsymbol{0}_{K\times n}]\\
		\end{aligned}
	\end{equation}

	For notational convenience, we use $\varPhi_k$ in lieu of $\varPhi (s_k)$ in the rest of this paper. 
	According to \eqref{eq:P2P_problem}, a quadratic problem (QP) can be established to realize P2P trajectory tracking. Consequently, the optimal problem is defined as follows.
	\begin{equation}\label{eq:MPC_ob_function}
		\begin{aligned}
			&\underset{\varDelta u_k ,\varepsilon}{\min}J\left( \varPhi_k ,\varDelta u_k ,\varepsilon \right)\\ &\;\;\;\;\;\;\;\;=\sum\limits_{i=1}^{N_p}{\lVert y_{k+i} -y ^{ref}_{k+i} \rVert _{Q}^{2}} 
			+\sum\limits_{i=0}^{N_c-1}{\lVert \varDelta u_{k+i} \rVert_R^2 +\rho \varepsilon ^2}
		\end{aligned}
	\end{equation}
	\begin{equation}
		\begin{aligned}
			\varDelta u_{\min}-\varepsilon \boldsymbol{1}_{m}<&\varDelta u_t <\varDelta u_{\max}+\varepsilon \boldsymbol{1}_{m}\\
			u_{\min}<&u_k <u_{\max}			
		\end{aligned}
	\end{equation}
	where $N_p$, $N_c$ indicate prediction and control horizons respectively. $\Delta u_k$ is the incremental control at $k$ time step, and $Q$ and $R$ are positive definite weight matrices while $\rho$ represents the penalty of the slack factor $\varepsilon$. 
	
	\begin{remark}
		Constraints on states are not added to avoid unsolvable situations leading to inconvenience on simulation performance comparison. Therefore, there could be results that have large offsets on poses and associated velocities if the identified model is not precise enough.
	\end{remark}
	
	As stated in \eqref{eq:MPC_dynamics}, prediction states in $N_p$ horizon, $\mathcal{Y}_k=[y_{k+1},\cdots ,\,y_{k+N_p}]^{\top}$, can be given as follows.
	\begin{equation}\label{eq:MPC_multistep_pred}
		\mathcal{Y}_k = \mathscr{A} \varPhi _k + \mathscr{B}\Delta U_k
	\end{equation}
	where
	\begin{equation}\nonumber
		\begin{aligned}
			\mathscr{A} &=[\mathbb{C} \mathbb{A} \,,\mathbb{C} \mathbb{A} ^2,\,\cdots ,\,\mathbb{C} \mathbb{A} ^{N_p}]^{\top}	\\
			\mathscr{B} &=\left[ \begin{matrix}
				\mathbb{C} \mathbb{A} ^0\mathbb{B}&		\boldsymbol{0}_{K\times n}&		\cdots&		\boldsymbol{0}_{K\times n}\\
				\mathbb{C} \mathbb{A} ^1\mathbb{B}&		\mathbb{C} \mathbb{A} ^0\mathbb{B}&		\cdots&		\boldsymbol{0}_{K\times n}\\
				\vdots&		\vdots&		\ddots&		\vdots\\
				\mathbb{C} \mathbb{A} ^{N_p-1}\mathbb{B}&		\mathbb{C} \mathbb{A} ^{N_p-2}\mathbb{B}&		\cdots&		\mathbb{C} \mathbb{A} ^{N_p-N_c}\mathbb{B}\\
			\end{matrix} \right]	\\
			\Delta U_k&=\left[ \Delta u_k,\cdots ,\Delta u_{k+N_c-1} \right]^{\top}.
		\end{aligned}
	\end{equation}

	In line with \eqref{eq:MPC_ob_function}, the objective function is reconstructed as follows.
	\begin{equation}\label{eq:final_ob_fun}
		\begin{aligned}
			\underset{\Delta U_k,\varepsilon}{\min}J& \left( \varPhi _k,\Delta U_k,\varepsilon \right) =\left[ 2\mathcal{E} _{k}^{\top}\mathcal{Q} \mathscr{B} \,\,\boldsymbol{0} \right] \left[ \begin{array}{c}
				\Delta U_k\\
				\varepsilon\\
			\end{array} \right] +\mathcal{E} _{k}^{\top}\mathcal{Q} \mathcal{E} _k
			\\
			&+\left[ \Delta U_{k}^{\top}\,\,\varepsilon \right] ^{\top}\left[ \begin{matrix}
				\mathscr{B} ^{\top}\mathcal{Q} \mathscr{B} +\mathcal{R}&		\boldsymbol{0}\\
				\boldsymbol{0}&		\rho\\
			\end{matrix} \right] \left[ \begin{array}{c}
				\Delta U_k\\
				\varepsilon\\
			\end{array} \right]
		\end{aligned} 
	\end{equation}
with the constraints
	\begin{equation}
		\begin{aligned}
			&\mathcal{Y} _k=\mathscr{A} \varPhi (s_k)+\mathscr{B} \Delta U_k,\ \  U_{\min}<U<U_{\max},
			\\
			&\varDelta U_{\min}<\varDelta U_k<\varDelta U_{\max}, \ \ \varepsilon _{\min}<\epsilon <\varepsilon _{\max}
		\end{aligned}
	\end{equation}
	where $\mathcal{E} _k=\mathscr{A}\varPhi_k - \mathcal{Y}^{ref}_k$, $\mathcal{Y} _{k}^{ref}$ is the reference sequence of states which has the same size with $\mathcal{Y}$, $U_k=[u_{k}, \cdots, u_{k+N_c-1}]^{\top}$ is the control sequence in prediction horizon, $\mathcal{Q}=\text{diag}([Q^{(1)}, \cdots, Q^{(N_p)}])$, $\mathcal{R}=\text{diag}([R^{(1)}, \cdots, R^{(N_c)}])$, and $^{(*)}$ denotes the $*$-th element of the diagonal matrix. Note that the term $\mathcal{E} _{k}^{\top}\mathcal{Q} \mathcal{E} _k$ is ignored because it is irrelevant with $\Delta U_k$ and $\varepsilon$. \par
	
	Through solving the QP in \eqref{eq:final_ob_fun}, the final control sequence $U_k$ can be obtained based on $\Delta U_k$. And the first control of $U_k$ is applied to the CarSim environment each time step. The implementation of DDK-MPC is shown in Algorithm \ref{alg:DDK-MPC}.
\begin{algorithm}
  \caption{DDK-MPC}\label{alg:DDK-MPC}
  \begin{algorithmic}[1]
	\Require The trained $\theta_e$, $\Lambda$, $B$; Initialize $N_p$, $N_c$, $Q$, $R$, $\rho$;
	\State Structure $A$ based on $\Lambda$ with \eqref{eq:A}.
    \For{\texttt{k=1, 2, ...}}
    	\State Transform current state and the reference trajectory to vehicle coordinate and obtain corresponding latent states, $\varPhi_k$ and $\mathcal{Y}^{ref}_k$, with \eqref{eq:koopman_eigenfunctions}.
      	\State Calculate the predicted latent states $\mathcal{Y}$ in $N_p$ horizon with \eqref{eq:MPC_multistep_pred}.
      \State Solve (\ref{eq:final_ob_fun}) to acquire a sequence of optimal control increments $\Delta U_k$.
      \State Get the optimal control $U_k$ and apply $U^{(1)}_k$ to CarSim.
    \EndFor
  \end{algorithmic}
\end{algorithm}

\section{Simulation validation}\label{sec:simulations}
CarSim is a high-fidelity vehicle dynamics simulation environment for vehicle control validation, even for drifting control\cite{zhang2017autonomous}. In this section, CarSim provides training, validation, and testing datasets. DDK is validated in two aspects, i.e., multi-step prediction and P2P trajectory tracking control combining Carsim.
\subsection{Datasets collection and preprocessing}
The datasets in \cite{xiao2020deep} which consist of 40 episodes are utilized in this paper. We randomly select 30 episodes for the training dataset, 5 episodes for the validation dataset, and the rest 5 episodes for the testing dataset. By combining CarSim 2019 with MATLAB/Simulink, datasets are collected from a C-CLass sedan car under sequences of control with driving force steering wheels and pedals of the Logitech G29. Each episode comprises $1000\sim 4000$ time steps with the sampling time of $10ms$. For the purpose to control the vehicle dynamics practically, some constraints are imposed, which are the steering wheel angle (SWA) $\zeta \in \left[ -7.85, 7.85\right] rad$, and the engine $\eta$ which consists of the brake pressure and throttle opening in the range of $[0, 10]MPa$ and $[0, 0.2]$. Data of velocities in collected datasets have a bound, that is $v_x \in [0, 27]\ m/s$, $v_y \in [-1.4, 1.7]\ m/s$, and $\dot{\psi}\in [-1.1, 1.1]\ rad/s$. Note that positive values of the engine denote the throttle opening while negative values indicate the brake pressure vice versa. 

\begin{figure}[htbp]
	\setlength{\abovecaptionskip}{-0.1cm}
	\setlength{\belowcaptionskip}{-1cm}
	\begin{center}
		\includegraphics[scale=0.5]{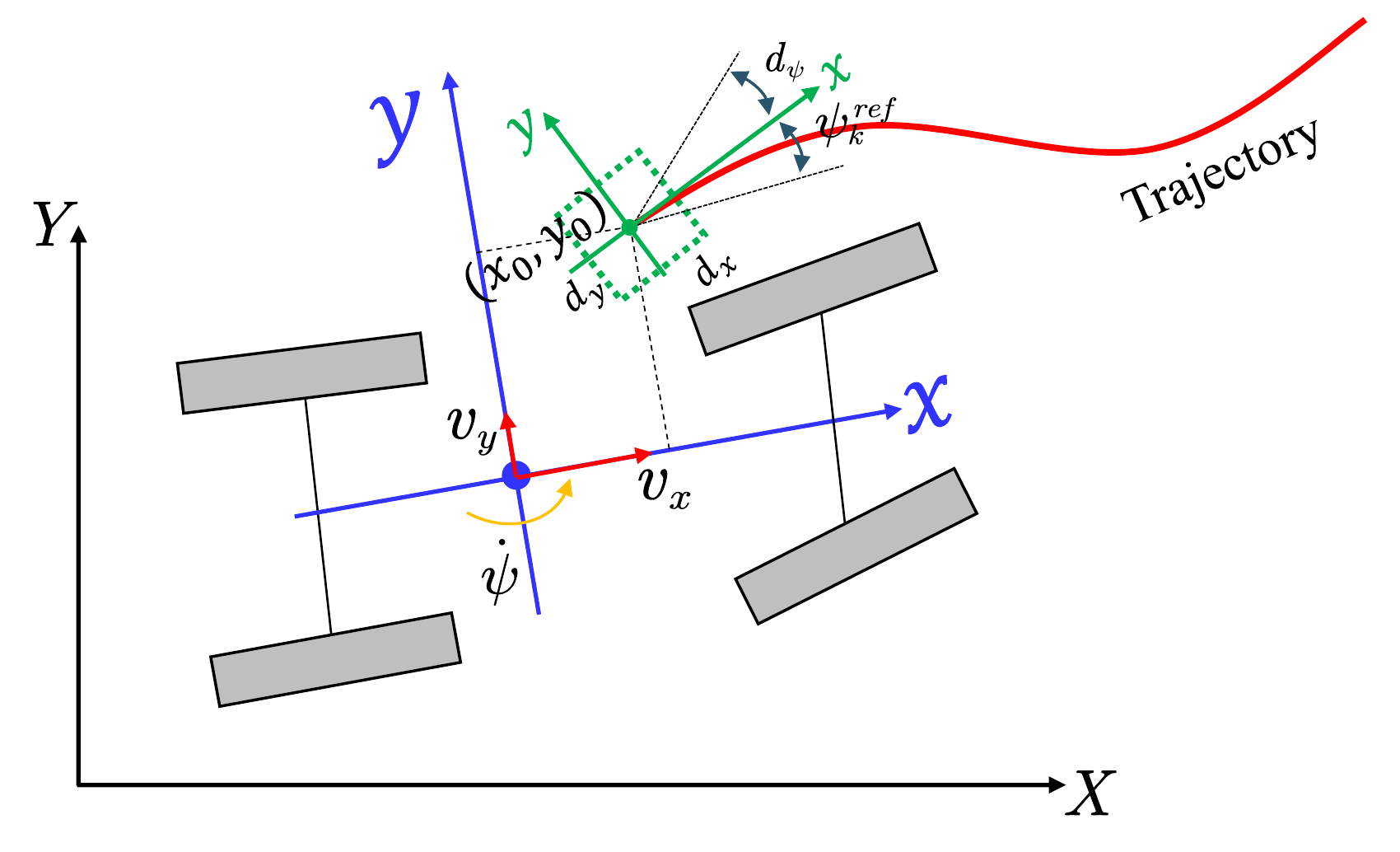}
		\caption{Coordinate transformation. The red solid line denotes a trajectory for training or tracking. The black coordinate signifies the global coordinate of the CarSim simulation environment. Trajectories are transformed from the global coordinate to vehicle coordinates by randomly selecting a point in the range of $[-d_x, d_x]$, $[-d_y, d_y]$, and $[-d_{\psi}, d_{\psi}]$ as the origin (the green coordinate) in the training process for improving the robustness while the vehicle's center of gravity (COG) plays the role of the coordinate transformation origin in controlling. Note that coordinate transformation appears at every training and controlling step, and only the pose data is transformed.}
		\label{fig:coor_transform}
	\end{center}
\end{figure}

The preprocessing contains two stages. The first stage happens before training which copes with variables that are irrelevant to the coordinate including the longitudinal and lateral velocities, yaw rate, steering wheel angle, throttle opening, and brake pressure. In the first stage, velocity variables are normalized to the range of $[-2, 2]$ while the control variables are transformed to the range of $[-1, 1]$. The second stage deals with pose variables in sampling process because they are constantly changing according to the origin of vehicle coordinates. Pose sequences are transformed to the vehicle coordinate as shown in Fig. \ref{fig:coor_transform}, then normalize them to the range of $[-2, 2]$. Coordinate transformations are different between the training and controlling process. In the controlling process, the exact current pose of the ego vehicle can be gained but it is inaccessible in the training process. Therefore, the transform origin $(x_o, y_o, \psi_o)$ is randomly selected with three uniform distribution, $x_o \sim U(-d_x, d_x)$, $y_o \sim U(-d_y, d_y)$, and $\psi_o \sim U(-d_{\psi}, d_{\psi})$ at each training step for improving the robustness.

Main hyperparameters are outlined in Table \ref{tab:training_hyperparameters}. The dimension of the latent state $K$ is an important hyperparameter that too small values lead to poor identification performance, but too big values cost a large calculation resulting in the trouble of real-time in control on the contrary. After $K$ is determined, the numbers of the Koopman eigenvalues are allocated as follows.
\begin{equation}\label{eq:num_eigevalues}
	\begin{aligned}
		N_{CP}&=integer\left( K/2 \right) 		\\
		N_R&=K\%2
	\end{aligned}
\end{equation}
where $N_{CP}$ and $N_R$ denote the number of conjugate complex pairs and real eigenvalues respectively, and $integer$ indicates the operator that obtains the minimum integer.

The encoder and decoder are realized with two MLPs that have the structure of $[mc, 64, 128, K, K]$ and $[K, 128, 64, 64, m]$, respectively. This work was trained by using the Python API in PyTorch-Lightning framework with an Adam optimizer based on an NVIDIA GeForce GTX 2080 Ti GPU. The corresponding MPC algorithm is realized in MATLAB/Simulink 2020a combining CarSim 2019 environment with an Intel i7-11700KF@3.6GHz.

\begin{table}
	\begin{center}
		\caption{Hyperparameters in DDK.}
		\scalebox{1.0}{
			\begin{tabular}{cccc}
				\toprule  
				\textbf{H-param} & \textbf{Value} & \textbf{H-param} & \textbf{Value}\\
				\midrule  
				Learning rate & $10^{-4}$ & Batch size & 256\\
				$K$ & 22 & $p$ & 50 \\
				$\alpha_1$ & 1.0 & $\alpha_2$ & 1.0 \\
				$\alpha_3$ & 1.0 & $\alpha_4$ & $10^{-6}$ \\
				$d_x$	&	$2m$ & $d_y$	&	$2m$	\\
				$d_{\psi}$ &	$20\degree$ & $c$ & 1  \\
				\bottomrule 
				\label{tab:training_hyperparameters}
		\end{tabular}}
	\end{center}
\end{table}

\subsection{Performance validation}
In this subsection, the DDK and DDK-MPC are verified separately. DeepEDMD \cite{xiao2020deep} and LRAN \cite{otto2019linearly} are Koopman operator-based methods with deep learning and obtain much better performance than kernel-based DMD and EDMD for approximating dynamics. In addition, a weak version of DDK (WDDK) is also taken into consideration. Except for prediction validation, MPCs of DeepEDMD, LRAN, DDK, and WDDK are adopted to realize trajectory tracking and the results can verify the advantages of concatenating the original state. In addition,  LTV-MPC\cite{falcone2008linear} and PurePursuit are also taken into comparison in the trajectory tracking process. 

\begin{remark}
	LRAN and WDDK are week versions of DeepEDMD and DDK which do not concatenate the original state with outputs of encoders as latent states. 
\end{remark}

\textbf{Validation of prediction:}
As outlined in Table \ref{tab:prediction_RMSE}, root mean square error (RMSE) of $120$ prediction steps is adopted to evaluate the prediction performance of different methods. Each method predicts and calculates its RMSE of prediciton according to the same episodes. And the results show that DDK has better capacities than DeepEDMD for approximating the vehicle dynamics. DDK and DeepEDMD overwhelm WDDK and LRAN in position verify that concatenating the original state can obtain better performance.
\begin{table}
	\caption{RMSEs of $120$ prediction steps for different methods}
	\centering
	\scalebox{.85}{
	\begin{tabular}{ccccccc}
		\toprule
		Alg & \makecell[c]{$x$\\ $(m)$} & \makecell[c]{$y$\\ $(m)$} & \makecell[c]{$\psi$\\ $(rad)$} & \makecell[c]{$v_x$\\ $(m/s)$} & \makecell[c]{$v_y$\\ $(m/s)$} & \makecell[c]{$\dot{\psi}$\\ $(rad/s)$} \\
		\midrule
		LRAN & 0.916 & 1.800 & 0.062 & 2.400 & 0.099 & 0.044 \\
		DeepEDMD & 0.141 & 0.102 & 0.016 & 0.487 & 0.047 & 0.025\\
		WDDK & 0.296 & 0.267 & 0.012 & 0.334 & 0.048 & 0.024	\\
		DDK & 0.090 & 0.112 & 0.007 & 0.321 & 0.046 & 0.024 \\
		\bottomrule 
		\label{tab:prediction_RMSE}
	\end{tabular}}
\end{table}
\par
\textbf{Validation of P2P trajectory trakcing:} To further validate DDK, MPCs of LRAN, DeepEDMD, DDK, and WDDK, which are notated as $\star$-MPC, are applied to realize P2P trajectory tracking based on Simulink/CarSim environment. Because latent states of LRAN and WDDK do not correspond to physical meaning, diagonal penalty matrices of states and controls for WDDK-MPC and LRAN-MPC are chosen as $Q=\text{diag}([q_1, ...,q_K])$,$R=\text{diag}([r_1,...,r_n])$ where $q=1000$, $r=5$, while we fix $R$ and try different $Q$ and choose the best $Q$ for other methods. For LTV-MPC, it has the same parameter tuning way with DDK-MPC and the best performance is chosen. PurePursuit \cite{snider2009automatic} tracks trajectories with a constant longitudinal velocity, and the preview distance equals $k_d v_x$, where $k_d=0.2, v_x=7m/s$. Note that poses tracking is completed in the vehicle coordinate even though it is visualized in the global coordinate for convenience. 

$\mathbb{C}$ in \eqref{eq:MPC_dynamics} equals $[I_K,\boldsymbol{0}_{K\times n}]$ for LRAN-MPC and WDDK-MPC, and equals $[\text{diag}([I_{m}, \boldsymbol{0}_{K-m}]), \boldsymbol{0}_{K\times n}]$ for DDK-MPC and DeepEDMD-MPC respectively. Other hyperparameters are shown in Table \ref{tab:MPC_hyperparameters}. It is worth noting that the engine is piecewise to represent the brake pressure and throttle opening according to the sign then anti-normalize them to the original range for applying to the CarSim car.

\begin{table}
	\begin{center}
		\caption{MPC hyperparameters in the simulations.}
		\scalebox{1.}{
			\begin{tabular}{cccc}
				\toprule  
				\textbf{H-param} & \textbf{Value} & \textbf{H-param} & \textbf{Value} \\
				\midrule  
				$\varepsilon_{min}$ & 0 & $\varepsilon_{max}$ & 100 \\
				$u_{min}$ & $[-1.0, -1.0]^{\top}$ & $u_{max}$ & $[1.0, 1.0]^{\top}$ \\
				$\Delta u_{min}$ & $[-0.5, -0.5]^{\top}$ & $\Delta u_{max}$ & $[0.5, 0.5]^{\top}$ \\
				$\rho$ & 10 & $t_s$ & $10ms$  \\
				\bottomrule 
				\label{tab:MPC_hyperparameters}
		\end{tabular}}
	\end{center}
\end{table}

\begin{figure}[htbp]
	\setlength{\abovecaptionskip}{-0.1cm}
	\setlength{\belowcaptionskip}{-0.1cm}
	\begin{center}
		\includegraphics[scale=0.48]{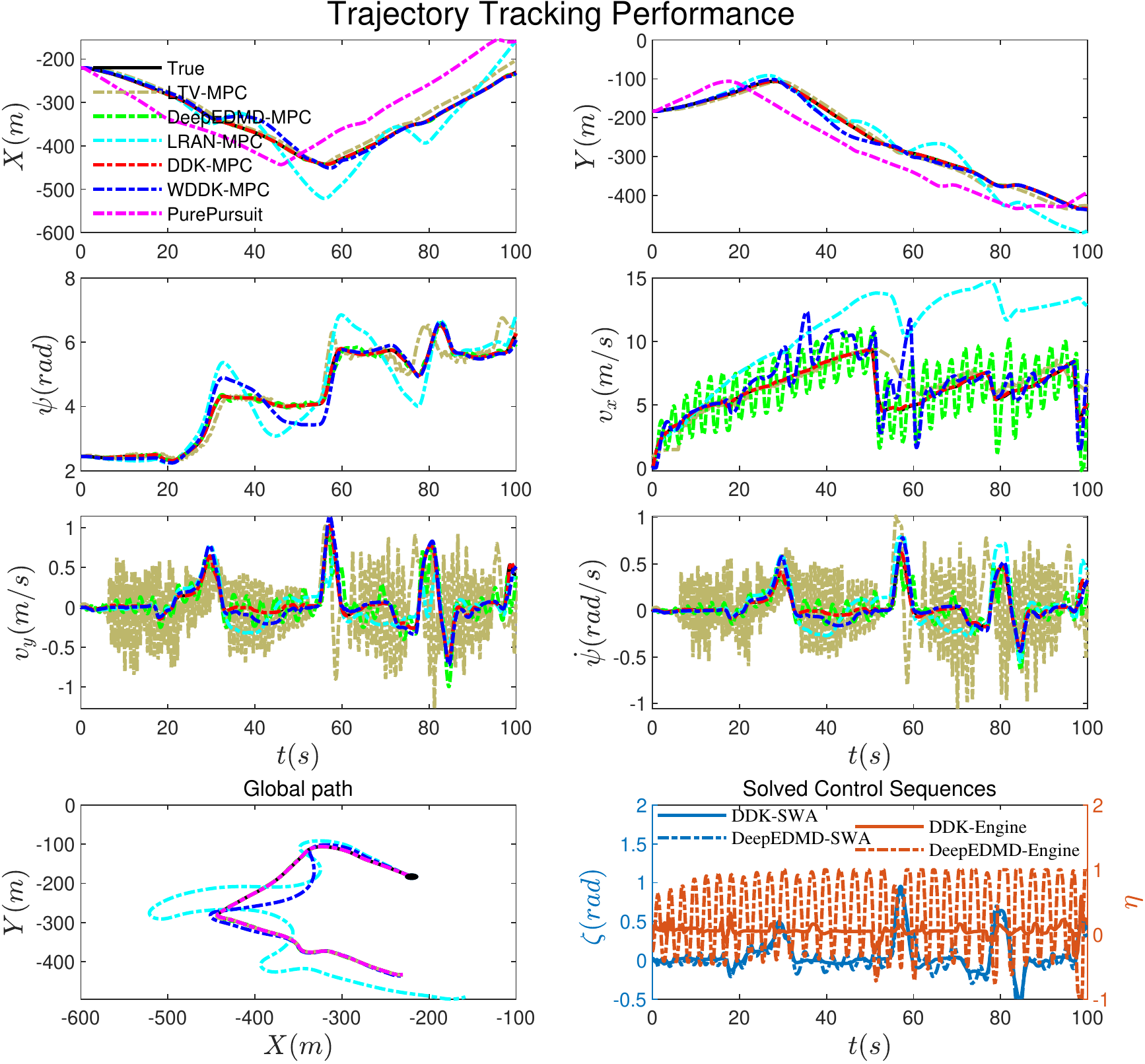}
		\caption{P2P trajectory tracking results for cases that prediction and control horizons equal $30$. The black dot in the bottom left subfigure is the start point.}
		\label{fig:mpc_control1}
	\end{center}
\end{figure}

\begin{figure}[htbp]
	\setlength{\abovecaptionskip}{-0.1cm}
	\setlength{\belowcaptionskip}{-0.1cm}
	\begin{center}
		\includegraphics[scale=0.52]{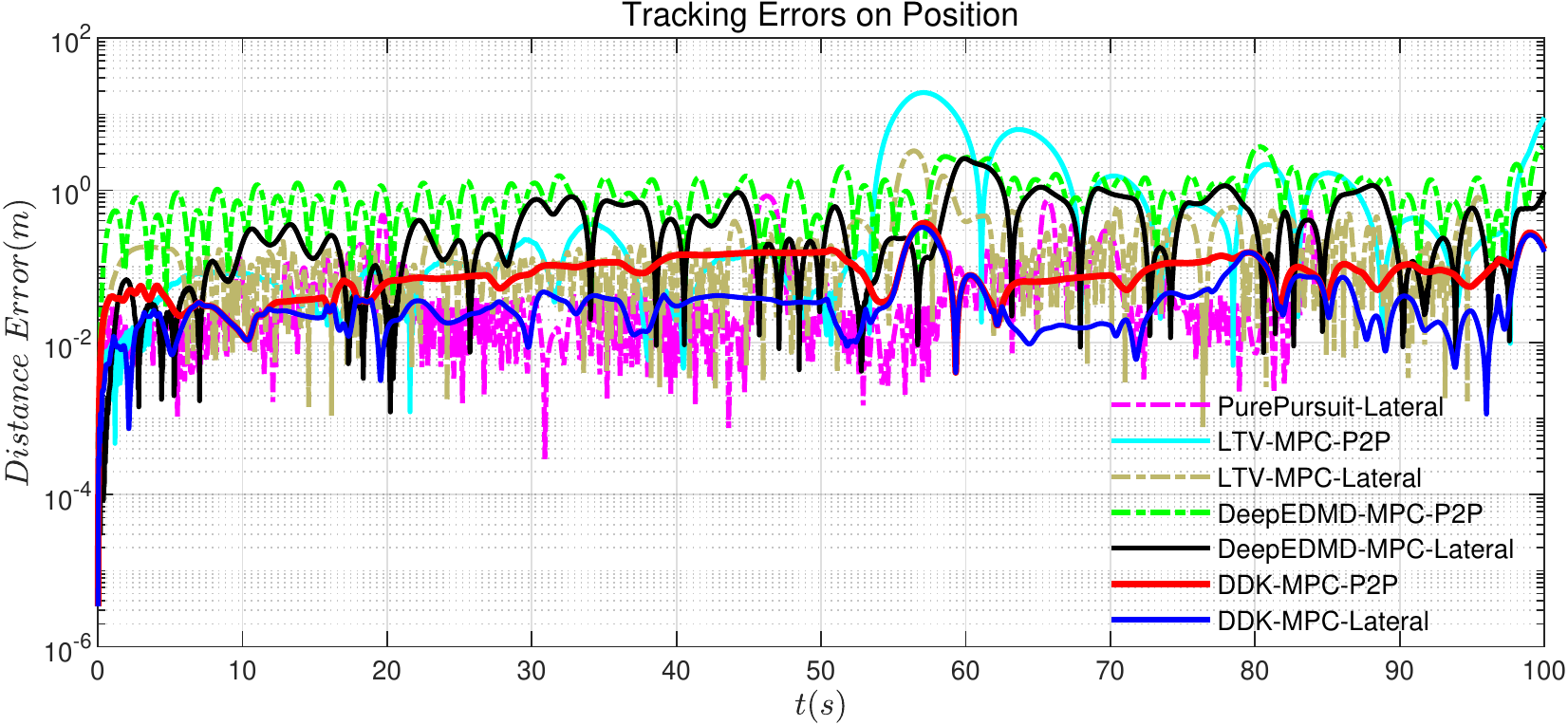}
		\caption{Tracking errors on position with different MPC methods for the trajectory in Fig. \ref{fig:mpc_control1}. $\star$-P2P denotes the P2P tracking errors whilst $\star$-Lateral denotes the tracking errors that equal the distance to the nearest point of the reference trajectory. In lateral tracking process, the reference a segment of trajectory started from the nearest point to the vehicle.}
		\label{fig:distance_error}
	\end{center}
\end{figure}

As shown in Fig. \ref{fig:mpc_control1}, LRAN-MPC and WDDK-MPC failed to track at the first bend. On the contrary, DDK-MPC and DeepEDMD track the trajectory successfully all the time. For clear visualization, we do not draw P2P and lateral tracking errors of WDDK-MPC and LRAN-MPC in Fig. \ref{fig:distance_error}. It is clear that algorithms with better interpretability gain much better performance, i.e. DDK-MPC and DeepEDMD-MPC. PurePursuit has excellent lateral tracking results with maximum and mean lateral errors equal $0.84m$ and $0.07m$ respectively, but it can not realize P2P tracking and it usually tracks with a constant velocity. 

\begin{figure}[htb]
	\setlength{\abovecaptionskip}{-0.1cm}
	\setlength{\belowcaptionskip}{-1cm}
	\begin{center}
		\includegraphics[scale=0.5]{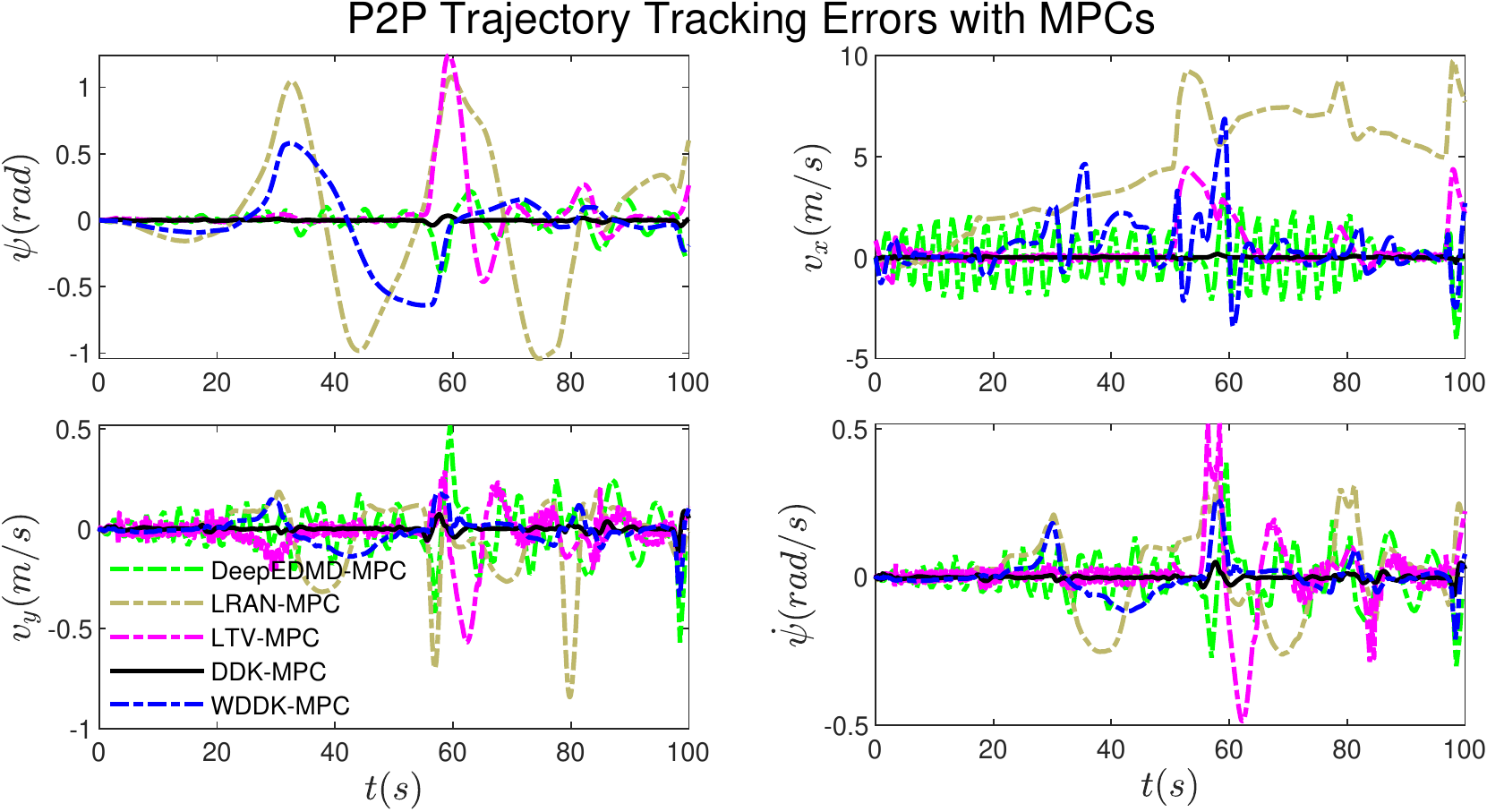}
		\caption{P2P tracking errors on yaw and velocities for the trajectory in Fig. \ref{fig:mpc_control1}.}
		\label{fig:mpc_control_error1}
	\end{center}
\end{figure}
P2P tracking errors on yaw angle and associated velocities are drawn in Fig. \ref{fig:mpc_control_error1} and detailed in Table \ref{tab:control_errors} of which demonstrate AME of DDK-MPC is smaller than $10cm$. The maximum offset equals $37cm$ and it occurs at the second bend. LTV-MPC has a good tracking capacity in straight roads for lateral tracking, but it performs terribly for P2P tracking. In addition, LTV-MPC costs a lot of time because it needs to calculate system matrices at each time step. The details of maximum and mean absolute tracking errors for DDK-MPC, DeepEDMD-MPC, and LTV-MPC with two different prediction and control horizons are shown in Table \ref{tab:control_errors}. 

Due to the large deceleration and yaw acceleration of the second bend, almost all methods have a large tracking error here. Especially for LTV-MPC, it tracks other road segments with a high precision except for the second bend. P2P trajectory tracking is more difficult because P2P trajectory tracking does not update the reference trajectory according to the nearest point so that it is hard to restore following while the vehicle has already deviated from the lane. The performance of DDK-MPC on P2P trajectory tracking validate the feasibility of DDK and its application value for control.

\begin{table}
	\caption{{Maximum and mean absolute errors for P2P tracking}}
	\centering
	\scalebox{.7}{
		\begin{tabular}{cccccccccc}
			\toprule
			\multicolumn{2}{c}{$\star$-MPC} & $N_p|N_c$& \makecell[c]{$P2P$ \\ $(m)$} & \makecell[c]{$Lat$\\$(m)$} & \makecell[c]{$\psi$ \\ $(rad)$} & \makecell[c]{$v_x$ \\ $(m/s)$} & \makecell[c]{$v_y$ \\ $(m/s)$} & \makecell[c]{$\dot{\psi}$ \\ $(rad/s)$} & \makecell[c]{$T_s$ \\ $(ms)$} \\
			\midrule
			\multirow{6}{*}{\rotatebox{90}{Mean}}&\multirow{2}{*}{\textbf{DDK}} & $30|30$ & 0.09 & 0.04 & 0.005 & 0.02 & 0.009 & 0.006& 8.0\\
			& &$40|40$ & 0.09 & 0.04 & 0.004 & 0.02 & 0.008 & 0.006 & 9.7 \\	
			& \multirow{2}{*}{\makecell[c]{Deep\\EDMD}} & $30|30$ & 0.97 & 0.42 & 0.06 & 1.52 & 0.11 & 0.07 & 8.1 \\
			& & $40|40$ & 0.94 & 0.53 & 0.07 & 1.04 & 0.10 & 0.06 & 10.3 \\
			& \multirow{2}{*}{\makecell[c]{LTV}} & $30|30$ & 2.12 & 0.24 & 0.12 & 0.56 & 0.06 & 0.05 & 19.1 \\
			& & $40|40$ & 2.08 & 0.4 & 0.13 & 0.60 & 0.08 & 0.07 & 30.7 \\
			\midrule
			\multirow{6}{*}{\rotatebox{90}{Maximum}}&\multirow{2}{*}{\textbf{DDK}} & $30|30$ & 0.37 & 0.33 & 0.04 & 0.26 & 0.11 & 0.06 & 8.0\\
			& & $40|40$ & 0.33 & 0.29 & 0.04 & 0.29 & 0.12 & 0.06 & 9.7 \\
			& \multirow{2}{*}{\makecell[c]{Deep\\EDMD}} & $30|30$ & 3.82 & 2.62 & 0.33 & 4.48 & 0.61 & 0.40 & 8.1 \\
			& & $40|40$ & 2.99 & 2.75 & 0.41 & 4.06 & 0.58 & 0.39 & 10.3 \\
			& \multirow{2}{*}{\makecell[c]{LTV}} & $30|30$ & 20.80 & 3.3 & 1.12 & 4.41 & 0.46 & 0.45 & 19.1 \\
			& & $40|40$ & 20.50 & 2.80 & 1.24 & 4.44 & 0.66 & 0.63 & 30.7 \\
			\bottomrule 
			\label{tab:control_errors}
	\end{tabular}}
\end{table}

\section{Experiment}\label{sec:experiment}
In this section, a real vehicle is utilized to collect datasets for validating the modeling capacity of DDK for vehicle dynamics. Datasets consist of 20 episodes with a distance of about 60km, including data on straight roads, U-turns, roundabouts, lane change, left and right turns. In experiments, except the dimension of latent states $K$ equals $16$, and steps of multi-step loss functions in \eqref{equ:loss_function} $p$ equals $80$, other hyper-parameters are the same as previous simulations.
\begin{figure}[htb]
	\setlength{\abovecaptionskip}{-0.1cm}
	\setlength{\belowcaptionskip}{-1cm}
	\begin{center}
		\includegraphics[scale=0.7]{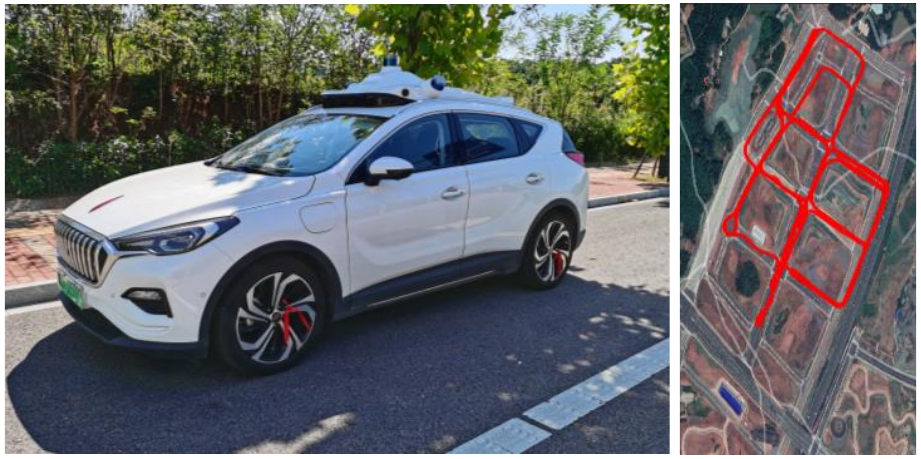}
		\caption{HQEHS3 SUV and one example episode. Differences with Carsim data include sampling interval is $20ms$ and the engine is represented by acceleration instead of the throttle and brake pressure.}
		\label{fig:HQEHS3}
	\end{center}
\end{figure}

As outlined in Table. \ref{tab:HQ_prediction_RMSE}, RMSEs of prediction with different prediction horizons are calculated by averages of 50 episodes. Prediction results demonstrate that DDK receives acceptable RMSE even for prediction more than 4s. The expressed performance is better than simulation and there are several possible reasons. One is that DDK adopts greater steps in loss functions so that it can predict longer for once at the latent state space. The other is that collected datasets with a real vehicle are more abundant on scenarios, e.g. roundabouts, U-turns, and et al. 

\begin{table}
	\caption{RMSEs of prediction with different horizons}
	\centering
	\scalebox{.85}{
		\begin{tabular}{ccccccc}
			\toprule
			\makecell[c]{Steps\\ $(20ms)$} & \makecell[c]{$x$\\ $(m)$} & \makecell[c]{$y$\\ $(m)$} & \makecell[c]{$\psi$\\ $(rad)$} & \makecell[c]{$v_x$\\ $(m/s)$} & \makecell[c]{$v_y$\\ $(m/s)$} & \makecell[c]{$\dot{\psi}$\\ $(rad/s)$} \\
			\midrule
			100(2s) & 0.178 & 0.168& 0.012 & 0.049 & 0.0195 & 0.007 \\
			200(4s) & 0.269 & 0.264 & 0.014 & 0.097 & 0.037 & 0.008	\\
			500(10s) & 0.353 & 0.353 & 0.019 & 0.203 & 0.079 & 0.014 \\
			\bottomrule 
			\label{tab:HQ_prediction_RMSE}
	\end{tabular}}
\end{table}

\begin{figure}[htb]
	\setlength{\abovecaptionskip}{-0.1cm}
	\setlength{\belowcaptionskip}{-1cm}
	\begin{center}
		\includegraphics[scale=0.49]{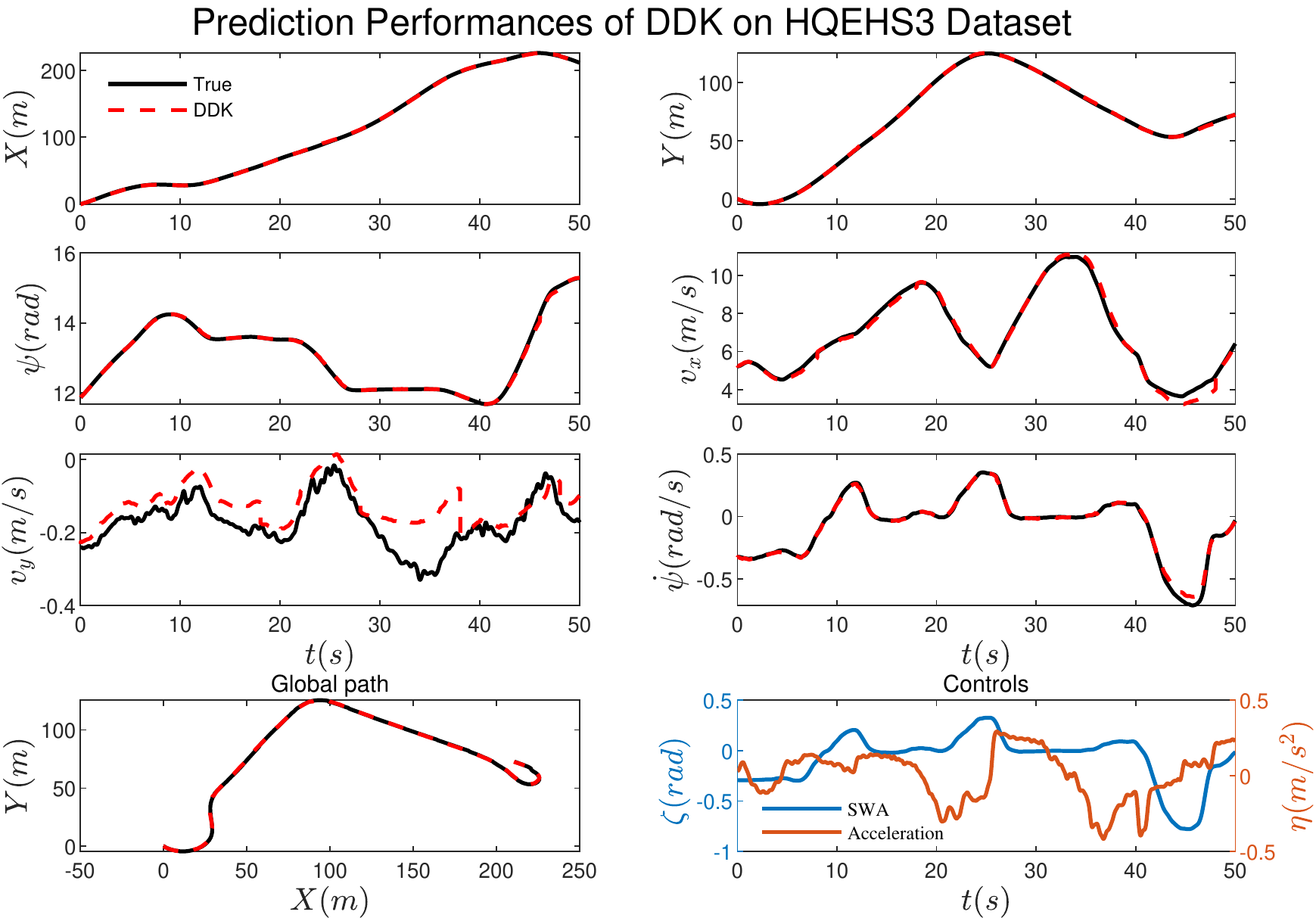}
		\caption{Prediction results of DDK on the dataset collected by HQEHS3 SUV. DDK predicts $500(10s)$ time steps each time. That is, the prediction sequence is composed of $5$ prediction sequences with a length of $10s$.}
		\label{fig:HQEHS3_DDK_prediction}
	\end{center}
\end{figure}

Fig. \ref{fig:HQEHS3_DDK_prediction} depicts a prediction sequence which is jointed with 5 prediction sequences with a length of $10s$. The sequence contains multiple types of roads, i.e. left and right turns, and a U-turn. Prediction results show an excellent prediction performance even there is a slightly large error on the longitudinal velocity.

\section{Conclusion}\label{sec:conclusion}
In this paper, a data-driven deep learning approach DDK is developed to approximate the Koopman operator of vehicle dynamics. The features of the proposed approach lie in the following three aspects. Firstly, the Koopman eigenvalues together with the system matrix B are learned so that DDK is also compatible with forced dynamics. Then, the physical properties of autonomous vehicles are matched by the former m elements of latent states corresponding to the identified state of the dynamics. Finally, an interpretable subsystem is extracted from the identified latent dynamics due to the diagonal characteristic of the state transition matrix. DDK-MPC is designed to realize P2P trajectory tracking based on a high-fidelity vehicle dynamics environment CarSim. Simulation results demonstrate excellent performances and validate the feasibility of DDK-MPC for P2P trajectory tracking under the condition of real-time requirements. The proposed approach is utilized to model the dynamics of a real electric SUV and the prediction results demonstrate the superiority of the proposed approach. Future research will involve trajectory tracking with learning-based methods to address the MPC parameter tuning problem, and DDK-MPC will be applied to realize P2P trajectory tracking on HQEHS3 in the near future. Furthermore, motion planning on scenarios that need P2P trajectory tracking will be studied.

\bibliographystyle{IEEEtran}
\bibliography{DeepKoopman}

\end{document}